\RequirePackage{ifpdf}
\documentclass[hyper]{JHEP3} 
\usepackage[utf8]{inputenc}
\pdfoutput=1
\usepackage{amssymb,amsmath,mathrsfs}
\usepackage{graphicx,rotate}
\usepackage{cite}

\usepackage{tikz}
\usepackage{multirow}
\usepackage{slashed}
\usepackage{color}
\let\normalcolor\relax
\usepackage{subfig}

\newcommand{\newbrace}[1][]{
\begin{tikzpicture}[baseline=-0.5ex]
\draw[#1] (0,0) -- (0.3,0.3);
\draw[#1] (0,0) -- (0.3,-0.3);
\end{tikzpicture}
}

{\;\newbrace[#1]\;\begin{array}{@{}l@{}}}%
{\end{array}}

\allowdisplaybreaks

\title{Constraints on Leptoquark Models from IceCube Data}

\author{Ujjal Kumar Dey\\
Physical Research Laboratory, Ahmedabad, Gujarat - 380 009, India \\
E-mail: \email{ujjaldey@prl.res.in}
}

\author{Subhendra Mohanty\\
Physical Research Laboratory, Ahmedabad, Gujarat - 380 009, India \\
E-mail: \email{mohanty@prl.res.in}
}

\abstract{Leptoquarks in the mass range of 500-1000 GeV can be resonantly produced in significant numbers by PeV neutrino  interacting with nuclei at IceCube. We compute the event rates of leptoquark production and decay events and use the 3-year IceCube data for PeV  energy events to find the allowed range of the leptoquarks mass and coupling parameter space. We use a low-scale quark lepton unification model based on the $SU(4)_C \otimes SU(2)_L\otimes U(1)_R$ gauge group where leptoquark couplings which give rise to proton decay are forbidden by the symmetry. We constrain the parameters of this model and point out signals of leptoquarks in this model which  may be seen in PeV energy IceCube events in the future.}


\begin{document}

\section{Introduction}
Leptoquarks (LQs) are $SU(3)$ coloured particles which also have non-zero lepton as well as baryon numbers. These arise in grand unified theories like  SO(10), SU(5) or Pati-Salam SU(4) and are expected to have masses in the GUT scale in order to not cause rapid proton decay. Models with low scale leptoquarks have been considered~\cite{Buchmuller:1986zs, Belyaev:2005ew, Dorsner:2005fq, Perez:2013osa} which do not lead to rapid proton  decays. The leptoquark model~\cite{Perez:2013osa} is based on the $G_{LQ}=SU(4)_C \otimes SU(2)_L\otimes U(1)_R$ quark-lepton unification group which breaks to the standard model at some low scale $\sim \rm TeV$. The SM quarks and leptons in this model can be unified in the same multiplet $(Q_L, \ell_L)\sim (4,2,0), (u_R,\nu_R) \sim ({\overline 4} ,1,-1/2)$ and 
$(d_R, e_R) \sim ({\overline 4} ,1,1/2)$. The scalars of this model give rise to this symmetry breaking and also to provide Yukawa couplings to generate fermion masses. They do not couple to the type of fermion-bilinears~\cite{Arnold:2013cva}($Q_L\cdot \ell_L, Q_L \cdot Q_L$ etc.) which give rise to proton decay. In addition to these scalar leptoquarks there are vector leptoquarks from the gauge bosons of the $G_{QL}$ gauge group but these are constrained to be heavy ($M_X>10^3 \,\rm TeV$) to evade bounds from  rare meson decays \cite{Valencia:1994cj, Smirnov:2007hv}. 

Leptoquark can be produced, singly or doubly, at colliders (hadron, $e^{+}e^{-}$ or $e^{\pm}p$) and signals (like the ${\ell \ell j j}$ from the decay of a leptoquark pair) and other important properties and constraints can be studied~\cite{Hewett:1987yg, Hewett:1989cs, Leurer:1993em, Ohnemus:1994xf, Choudhury:1994he, Hewett:1997ce, Abdullin:1999im, Kramer:2004df,Alikhanov:2012kk, Dorsner:2014axa, Mandal:2015vfa,Evans:2015ita}. 
A search for pair-production of first and second generation scalar LQs has been performed with 19.6 fb$^{-1}$ of data by CMS~\cite{CMS-PAS-EXO-12-041, CMS-PAS-EXO-12-042}. According to this the first generation scalar LQs with masses less than 1005 (845) GeV are excluded for $\beta = 1 (0.5)$, where $\beta$ represents the branching fraction of an LQ to a charged lepton and a quark. For second generation scalar LQs the exclusion limits are 1070 (785) GeV for $\beta = 1 (0.5)$. Similarly, scalar LQs of third generation with masses below 740 GeV are excluded at $95\%$ C.L.  assuming a $100\%$ branching fraction for the LQ decay to a tau lepton and a bottom quark~\cite{Khachatryan:2014ura}. A recent $2.6\sigma$ excess in a $\ell \ell jj\slashed{E}_{T}$ events, reported by CMS~\cite{CMS-PAS-SUS-12-019}, has been explained by various models of LQ in the mass range 550-650 GeV~\cite{Queiroz:2014pra, Allanach:2015ria}.

It has been pointed out~\cite{Anchordoqui:2006wc} that a natural arena for the production of leptoquarks is the  neutrino-nucleon interactions at IceCube~\cite{Aartsen:2014gkd}. Recently the leptoquark interaction of the form $S^\dagger (\ell_{L}.Q_{L} + u^{C} \tau^{C}) $ of a scalar leptoquark with the third generation leptons and first generation quarks has been proposed as an explanation of the 2 year IceCube data~\cite{Barger:2013pla}. A very recent account of a minimal leptoquark scenario in the light of CMS and IceCube observations are presented in~\cite{Dutta:2015dka}. Restriction to the third generation leptons explains the paucity of muon tracks at neutrino energies 0.1- 2 PeV and  the cross section is enhanced by the resonant production of $\sim 600$  GeV leptoquarks at neutrino energies $\sim \rm  PeV$.   

In this paper we discuss the production and decay of the scalar leptoquarks of the low scale quark-lepton unification model based on $SU(4)_C \otimes SU(2)_L\otimes U(1)_R$~\cite{Perez:2013osa} at IceCube. The leptoquarks  in this model decay into  hadronic showers (plus neutrinos). The analysis of the 988 days data gives a neutrino flavor ratio of 1:1:1 consistent with all 37 events being from standard neutrino CC and NC events~\cite{Aartsen:2015ivb}. Thus to put an upper bound on the parameters of the leptoquark model we make the conservative assumption that no event among the 35 with less than PeV energies are of leptoquark origin. The resonant production of leptoquarks with masses in the 500-1000 GeV, however, can give significant number of hadronic shower events at IceCube at PeV range. 

\section{Quark-Lepton Unification Model}

 We have used the low-scale quark-lepton unification model described in~\cite{Perez:2013osa} which predicts the existence of both vector and scalar leptoquarks . This leptoquark model is based on the $G_{QL}=SU(4)_c \otimes SU(2)_L\otimes U(1)_R$ quark-lepton unification group which spontaneously breaks to the standard model $G_{SM}=SU(3)\otimes SU(2)_L\otimes U(1)_Y$ at some low scale. The SM quarks and leptons can be unified in the same multiplet $(Q_L, \ell_L)\sim (4,2,0), (u_R,\nu_R) \sim ({\overline 4} ,1,-1/2)$ and 
$(d_R, e_R) \sim ({\overline 4} ,1,-1/2)$. 
We concentrate on the scalar LQs of this theory, namely, $\Phi_{3}\sim \left(\overline{3},2,-1/6\right)_{\rm SM}$ and $\Phi_{4}\sim \left(3,2,7/6\right)_{\rm SM}$ which are present in the scalar $\Phi$ of the theory. The masses of the components of $\Phi_{3}$ and $\Phi_{4}$ are  determined by the parameters of the scalar potential. The relevant part of the potential is given by~\cite{Perez:2013osa},
\begin{align}
\mathcal{V} =  m_{\Phi}^{2} & \text{Tr}[\Phi^{\dagger}\Phi] + 
     \tilde{\lambda}_{2}H^{\dagger}H\text{Tr}[\Phi^{\dagger}\Phi] + 
     \tilde{\lambda}_{3}\chi^{\dagger}\chi \text{Tr}[\Phi^{\dagger}\Phi] 
     + 
     \tilde{\lambda}_{5}H^{\dagger}\text{Tr}[\Phi^{\dagger}\Phi]H 
     \nonumber \\
     & \tilde{\lambda}_{6}\chi^{\dagger} \Phi \Phi^{\dagger} \chi + 
       \tilde{\lambda}_{9}\text{Tr}[\Phi^{\dagger}\Phi]^{2} + 
       \tilde{\lambda}_{10}(\text{Tr}[\Phi^{\dagger}\Phi])^{2},
\end{align}
where under $G_{LQ}$, $\Phi\sim (15,2,1/2)$, $H\sim (1,2,1/2)$ and $\chi\sim (4,1,1/2)$. Also,
\begin{align}
\Phi = \begin{pmatrix}
\Phi_{8} & \Phi_{4} \\
\Phi_{3} & 0
\end{pmatrix} + T_{4}H_{2},
\end{align}
where $T_{4}$ is the $SU(4)_{C}$ generator and $H_{2}$ is another Higgs doublet. Moreover the vacuum expectation values (vev) of the scalar fields are given by $\langle\chi^{0}\rangle = v_{\chi}/\sqrt{2}$, $\langle H^{0}\rangle = v_{1}/\sqrt{2}$ and $\langle H_{2}^{0}\rangle = v_{2}/\sqrt{2}$. With these vevs the mass terms for the fields $\Phi_{3,4,8}$ can be written as,
\begin{align}
\left(m_{\Phi}^{2}+\frac12\left((\tilde{\lambda}_{2}
       +\tilde{\lambda}_{5})v_{1}^{2} +
        \tilde{\lambda}_{3}v_{\chi}^{2}\right)\right)
       \left(\Phi_{3}^{\dagger}\Phi_{3} + 
        \Phi_{4}^{\dagger}\Phi_{4} + 
        \Phi_{8}^{\dagger}\Phi_{8}\right) + 
        \frac12 \tilde{\lambda}_{6}v_{\chi}^{2}
        \Phi_{3}^{\dagger}\Phi_{3},
\end{align}
From the mass terms it is evident that $\Phi_{8}$ and $\Phi_{4}$ are degenerate in mass but $\Phi_{3}$ mass can be different from the former two, owing to the $\tilde{\lambda}_{6}$ term. In this context it is worth mentioning that the LHC searches at $\sqrt{s}=8$ TeV with 19.7 fb$^{-1}$ of integrated luminosity, through a dijet resonance above 1.2 TeV have excluded the mass range of [1.3--2.5] TeV of the color-octet mass~\cite{Khachatryan:2015sja}; also according to a more recent search the bound on this mass exists down to 500 GeV~\cite{CMS:2015neg}. Thus the octet $\Phi_{8}$ and the leptoquark $\Phi_{4}$ can be well within this bound but the leptoquark $\Phi_{3}$ can be of smaller mass by taking small negative $\tilde{\lambda}_{6}$.

In this work we take them to be the free parameters which we constrain from the IceCube data. The relevant interactions of the leptoquarks are,
\begin{align}
\mathcal{L}_{\rm LQ} = \lambda_{2}Q_{L}\Phi_{3}\nu^{C}+
\lambda_{2}\ell_{L}\Phi_{4}u^{C}+
\lambda_{4}Q_{L}\Phi_{4}^{\dagger}e^{C}+
\lambda_{4}\ell_{L}\Phi_{3}^{\dagger}d^{C}+\rm{h.c.}
\label{e:lagrangian}
\end{align}

 The symmetries of LQs $\Phi_{3}$ and $\Phi_{4}$ do not allow interaction terms with type of fermion bilinears  ($Q_L \cdot \ell_L, Q_L\cdot Q_L$) that  lead to proton decay via these LQ exchanges at tree level at dimension four. However, there exists dimension five operators that may lead to proton decay. But these dimension five operators can be avoided by the imposition of appropriate discrete symmetries~\cite{Arnold:2013cva}. The constraints on these types of LQs coming from charged lepton sector, e.g., via the processes $\mu \to e\gamma$, conversion of $\mu$ to $e$ and electric dipole moment of the electron, have been discussed in~\cite{Arnold:2013cva}.

 The coupling $\lambda_{4} < 0.01$ and this constraint comes from the kaon decay $K_{L}^{0}\to e^{\mp}\mu^{\pm}$ to be in agreement with fermion masses~\cite{Smirnov:2007hv}. The coupling $\lambda_{2}$ is not  constrained and we put bounds on this coupling from IceCube events.  Constraints on leptoquark models from other rare decays has been considered in~\cite{Hiller:2014yaa, Varzielas:2015iva}.

\begin{table}[h]
\centering
\begin{tabular}{|c|c|c|}
\hline
Leptoquark & \begin{tabular}[c]{@{}c@{}}Electromagnetic Charge\\ (Q)\end{tabular} & Neutrino coupling \\ \hline \hline
\multirow{2}{*}{$\Phi_{3}$} & $+2/3$ & $\lambda_{2}\bar{\nu}u$ \\ \cline{2-3} 
 & $-1/3$ & $\lambda_{2}\bar{\nu}d$ \\ \hline
\multirow{2}{*}{$\Phi_{4}$} & $+5/3$ & -- \\ \cline{2-3} 
 & $+2/3$ & $\lambda_{2}\bar{\nu}u$ \\ \hline \hline
\end{tabular}
\caption{Leptoquarks with appropriate electromagnetic charges and couplings.}
\label{tab:1}
\end{table}

In Table~\ref{tab:1} we present the electric charges of the components of the LQs and their couplings with the neutrinos. Note that the state of $\Phi_{4}$ with electric charge $(+5/3)$ will have no coupling with neutrinos. All the processes relevant for the production of LQ from neutrino-nucleon interaction and its subsequent decay to  the $\tau$ and hadronic showers are (here the superscripts in $\Phi_{3,4}$ represents the corresponding electric charges.),

\begin{align}
\bar{\nu}u \to \Phi_{3}^{2/3} \to \bar{\nu}u, \quad
\bar{\nu}d \to \Phi_{3}^{-1/3} \to \bar{\nu}d,\quad
\bar{\nu}u \to \Phi_{4}^{2/3} \to \bar{\nu}u~,
\label{e:lqprocess} 
\end{align}
these processes are proportional to the coupling $\lambda_2$ on which the constraints are mild. 

On the other hand the model (\ref{e:lagrangian}) allows the following processes which could have contributed to the IceCube signal,
\begin{align}
\bar{\nu}u \to \Phi_{3}^{2/3} \to
\ell^{+}d, \quad
\bar{\nu}d \to \Phi_{3}^{-1/3} \to
\ell^{-}u ,\quad
\bar{\nu}u \to \Phi_{4}^{2/3} \to
\ell^{+}d 
\label{e2:lqprocess}
\end{align}
but these processes are proportional to
  the coupling $\lambda_{4}$ which is already constrained  to be $\lambda_4 < 0.01$ from kaon decay  $K_{L}^{0}\to e^{\mp}\mu^{\pm}$ ~\cite{Smirnov:2007hv}.  We study LQ production and decay processes where only the  $\lambda_{2}$ coupling arises.  Due to the smallness of $\lambda_4$ the LQ decays to $\ell^\pm +j $, which is  the signal in~\cite{Barger:2013pla}, is not significant in this model. We considered $\lambda_{2}$ to be $0.5$ and $1.0$ and we keep $\lambda_{4}$ to a fixed value 0.005 in our subsequent study.
  
The neutrino-nucleon cross section for the production and subsequent decay of LQ of mass $M_{\Phi}$, in the narrow width approximation, is given by~\cite{Doncheski:1997it, Alikhanov:2013fda},
\begin{align}
\sigma_{\rm LQ}(\bar{\nu}N\to \bar{\nu}X)=\frac{\pi}{2M_{\Phi}^{2}}\lambda_2^{2}~{\rm BR}(\Phi\to \bar{\nu}q)xq_{N}(x;\mu^{2}),
\label{e:signucl}
\end{align} 
where $q_{N}(x;\mu^{2})$ is the parton distribution function (PDF) of the parton $q$ in the nucleuon (with mass $m_{N}$) and $x={M_\Phi}^2/(2 m_N E_\nu)$ is the parton fractional momentum. The renormalization scale $\mu$ is set at $M_\Phi$. We have used the nuclear PDFs, \texttt{CTEQ6l1}~\cite{Pumplin:2002vw} at LO to get the $q_{N}(x;\mu^{2})$.

\begin{figure}[!htbp]
\centering
\subfloat[][]{
\includegraphics[scale=0.4]{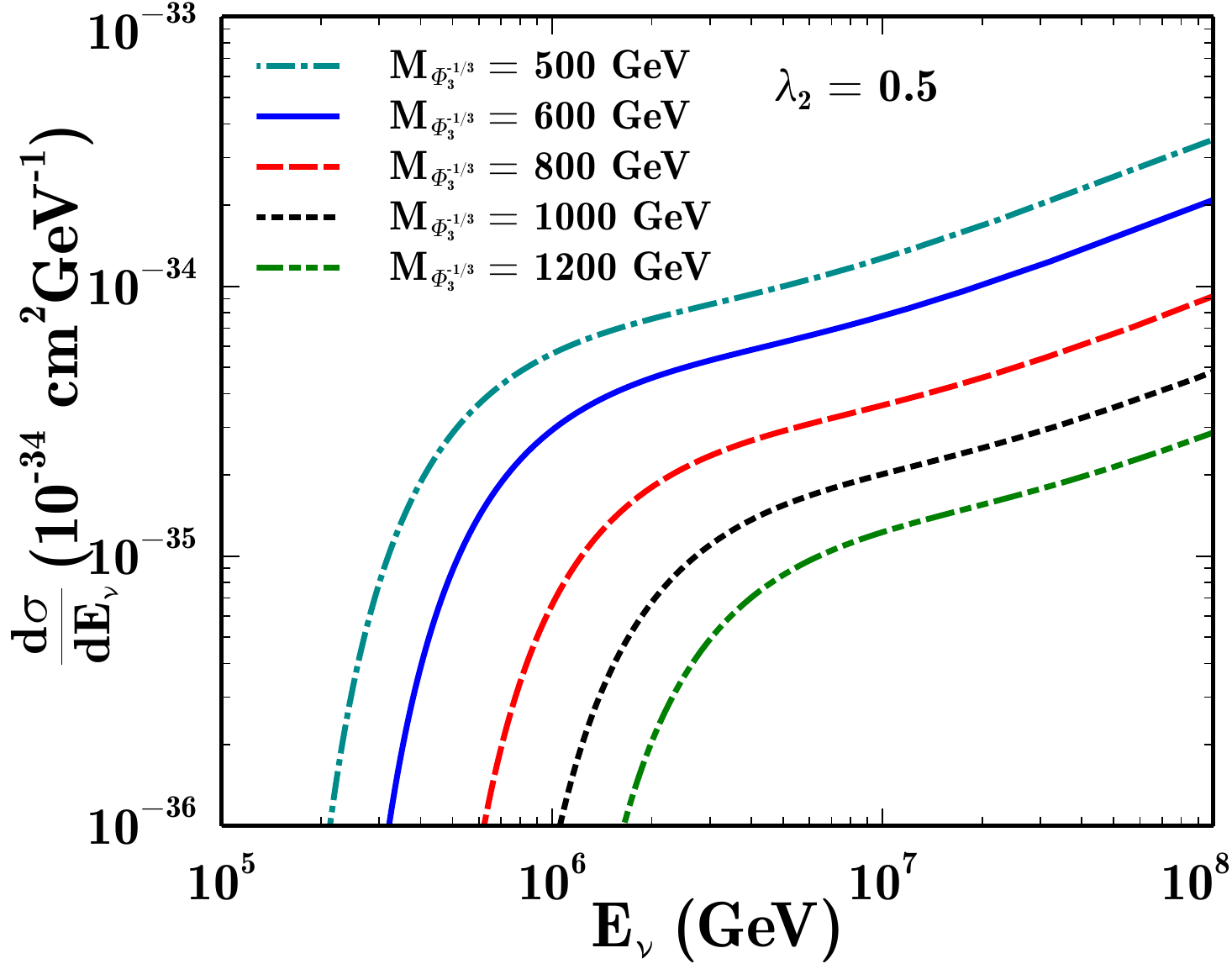}
\label{sf:dsigdEa}
}
~~~~~~~~
\subfloat[][]{
\includegraphics[scale=0.4]{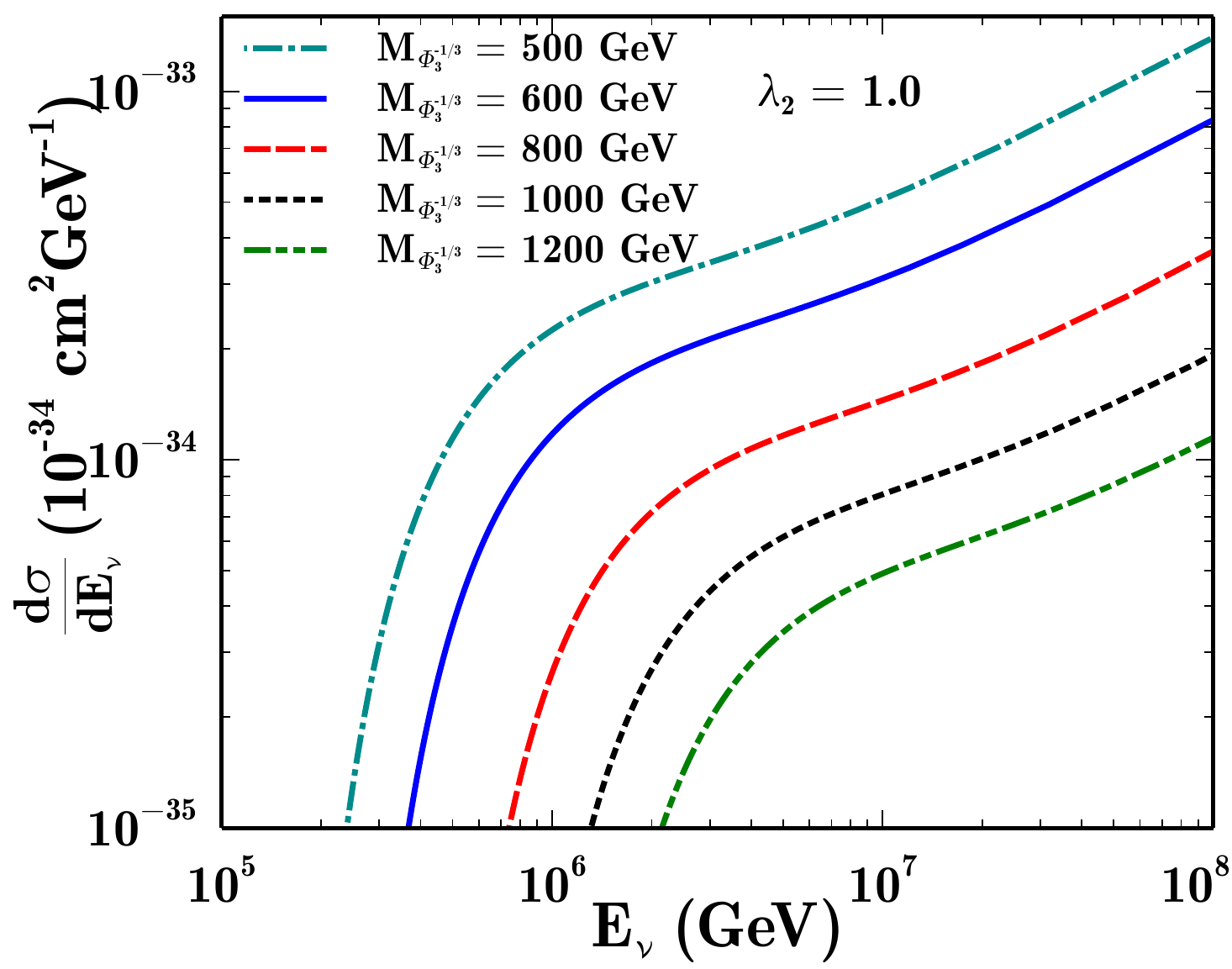}
\label{sf:dsigdEb}
}\\
\subfloat[][]{
\includegraphics[scale=0.4]{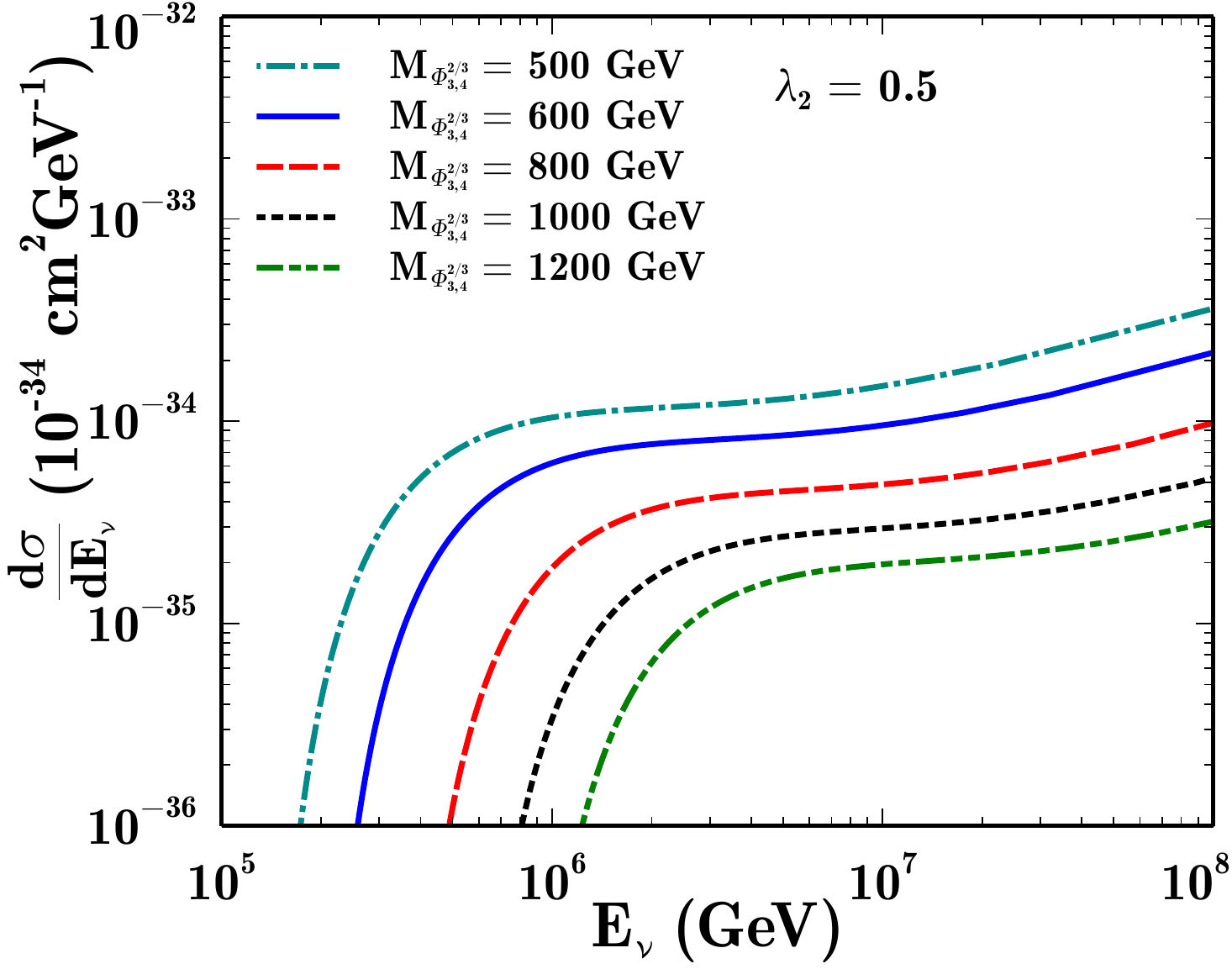}
\label{sf:dsigdEc}
}
~~~~~~~~
\subfloat[][]{
\includegraphics[scale=0.4]{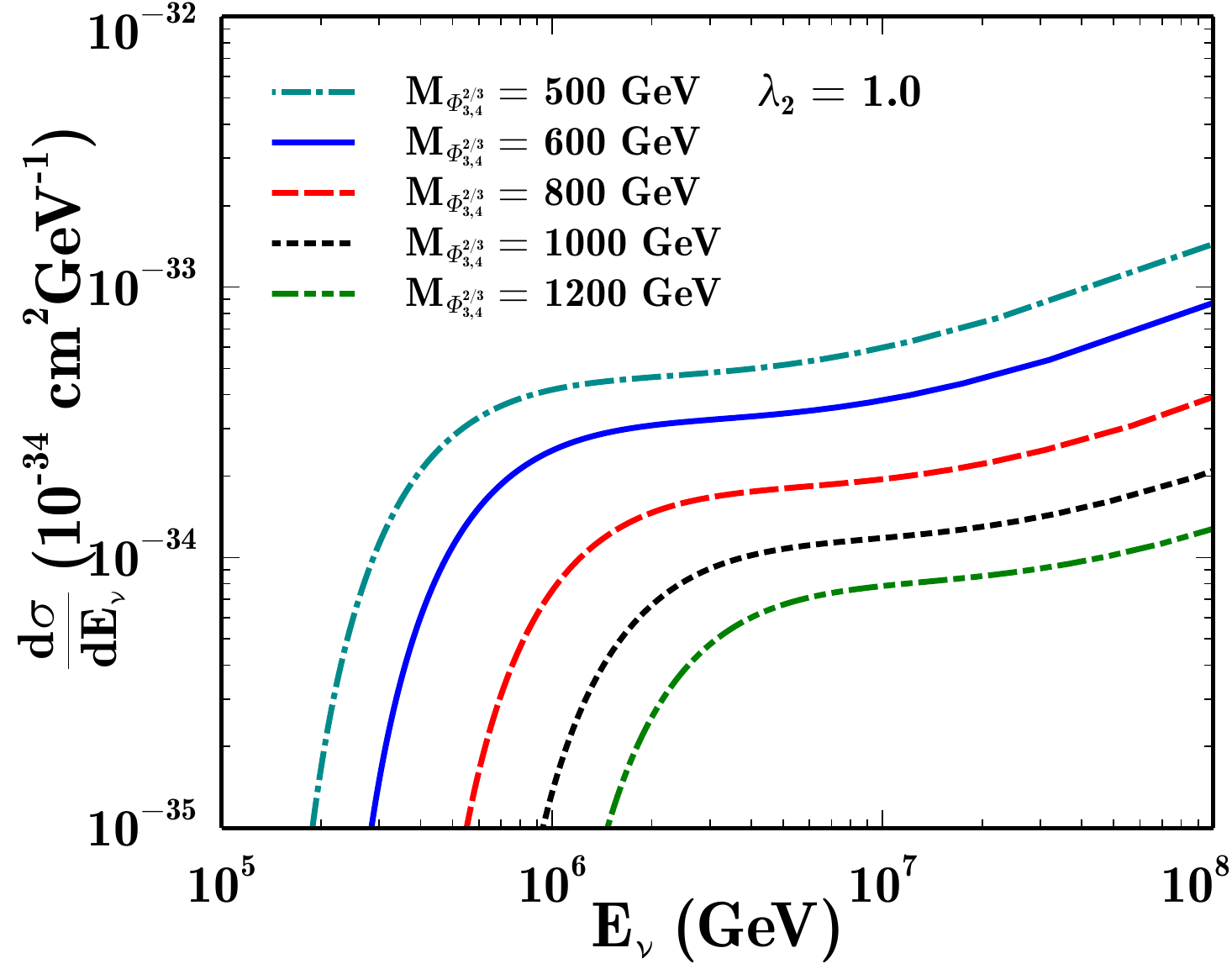}
\label{sf:dsigdEd}
}
\caption{The distribution of neutrino-nucleon cross section for various cases. The values of the coupling $\lambda_{2}$ that have been used are shown in each plot.}
\label{f:dsigdE}
\end{figure}
In Fig.~\ref{f:dsigdE} we show the distribution of neutrino-nucleon cross section, $\frac{d\sigma}{dE_{\nu}}(\bar{\nu}N\to \bar{\nu}X)$. The upper panel (Figs.~\ref{sf:dsigdEa} and~\ref{sf:dsigdEb}) shows the case where only the leptoquark $\Phi_{3}^{-1/3}$ is involved. In the lower panel (Figs.~\ref{sf:dsigdEc} and~\ref{sf:dsigdEd}) we show the distribution where either $\Phi_{3}^{2/3}$ or $\Phi_{4}^{2/3}$ comes in the intermediate state.

\section{Constraints from IceCube Events}
The three-year (2010-2012) data set, with a livetime of $988$ days, reveals $37$ neutrino events in the energy range 0.3-2 PeV. Among these, 9 are track events and 28 are cascade events, with the flavor ratio $1:1:1$~\cite{Aartsen:2015ivb} expected from pion/muon decay neutrinos oscillating over cosmological distances~\cite{Learned:1994wg, Beacom:2003nh}. The astrophysical neutrino flux (averaged over zenith angle) follows a power law~\cite{Aartsen:2014gkd},

\begin{align}
\Phi(E_{\nu}) = 4.74\times 10^{-7}\left(\frac{E_{\nu}}{1\rm GeV}\right)^{-2.3}{\rm GeVcm^{-2}s^{-1}sr^{-1}}.
\label{flux}
\end{align}

The highly energetic incoming neutrinos can produce leptoquarks after the collision with quarks in the nucleons at IceCube. The subsequent decay of the leptoquarks will be registered as events at IceCube. The expected number of events at IceCube is given by,

\begin{align}
\mathcal{N}_{LQ}=4\pi n_{T}T\int dE_{\nu}[\sigma_{\rm LQ}(E_{\nu}).{\rm Br}]\Phi(E_{\nu}),
\end{align}
where $n_{T}$ is the effective number of target nucleons in IceCube and is $\sim 6.0\times 10^{38}$ as the effective volume is roughly 1 km$^{3}$; the time of exposure $T=988$ days. 
$\sigma_{\rm LQ}(E_{\nu})$ has been defined in Eq.~\ref{e:signucl}; the ${\rm Br}$ represents the branching ratio of the leptoquark $\Phi$ decaying to a neutrino and a quark.

\begin{figure}[!htbp]
\centering
\subfloat[][]{
\includegraphics[scale=0.4]{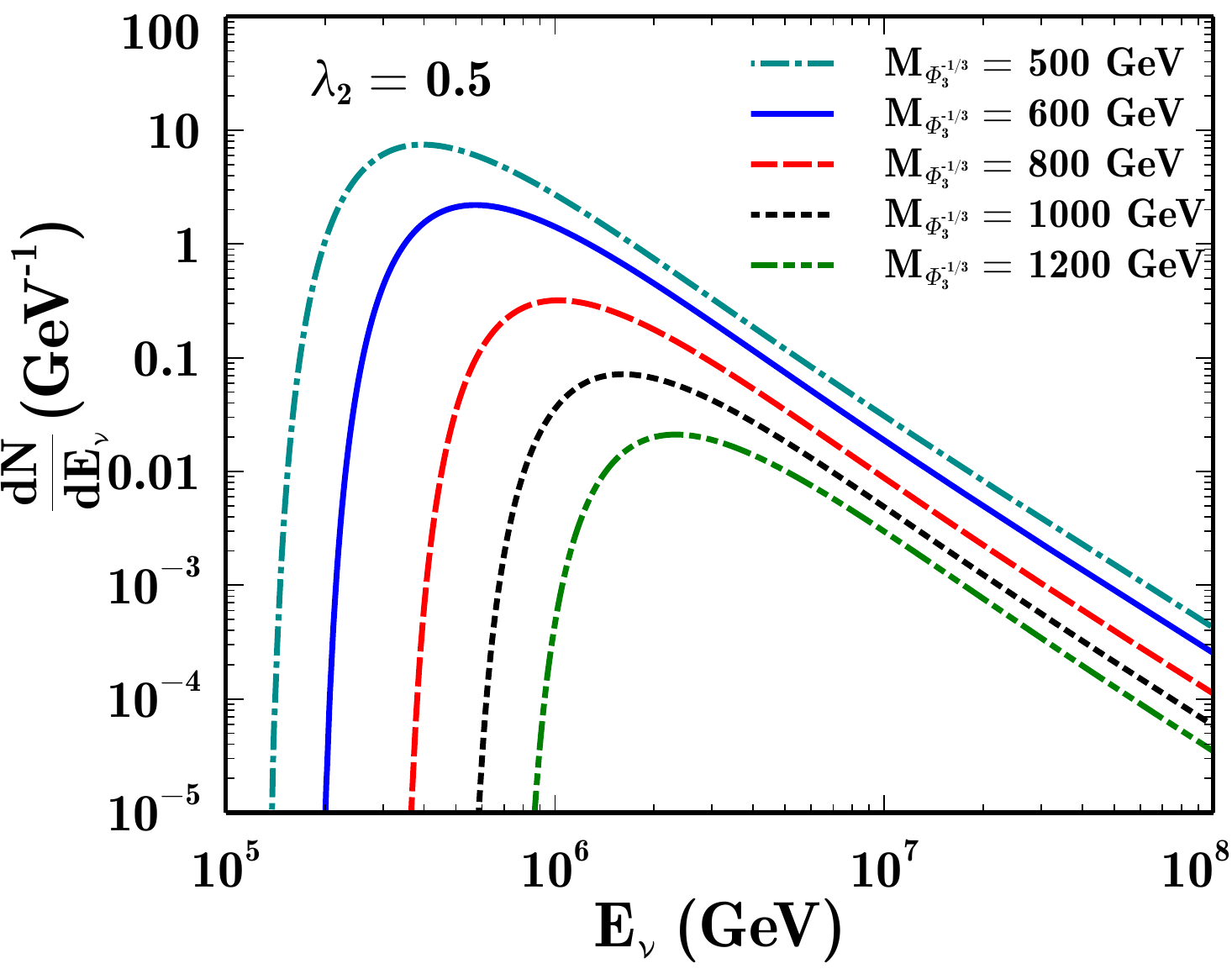}
}
~~~~~~~~
\subfloat[][]{
\includegraphics[scale=0.4]{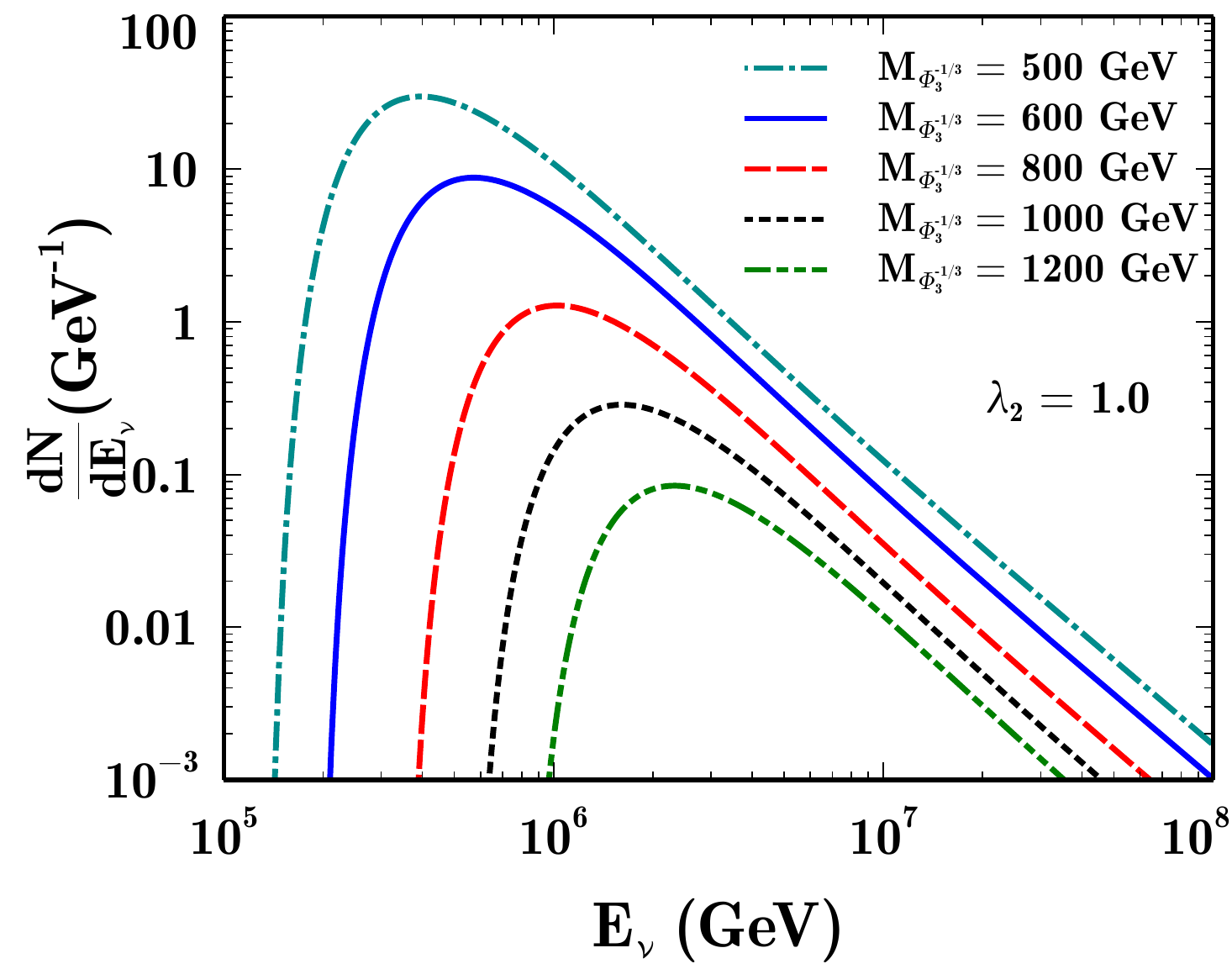}
}\\
\subfloat[][]{
\includegraphics[scale=0.4]{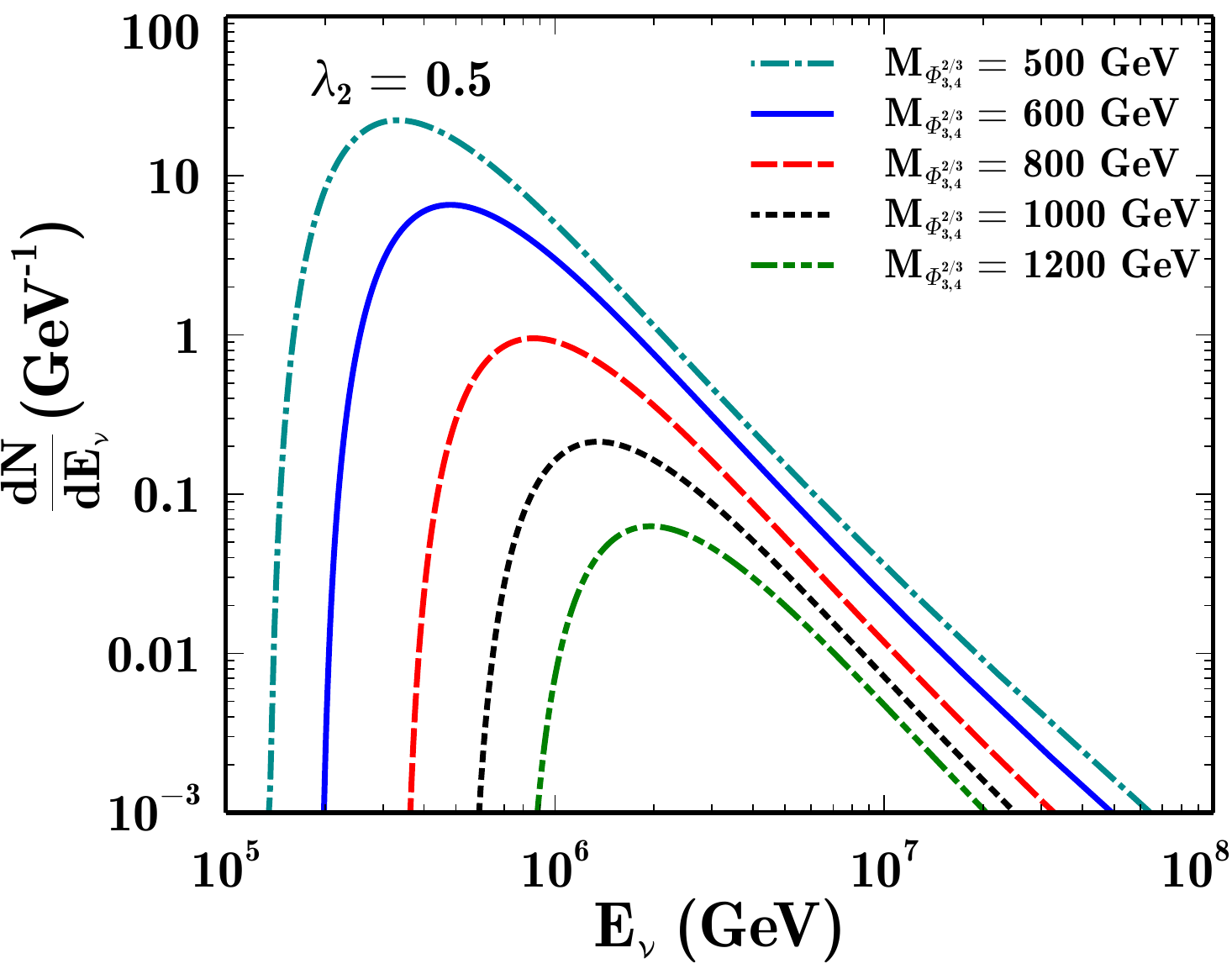}
}
~~~~~~~~
\subfloat[][]{
\includegraphics[scale=0.4]{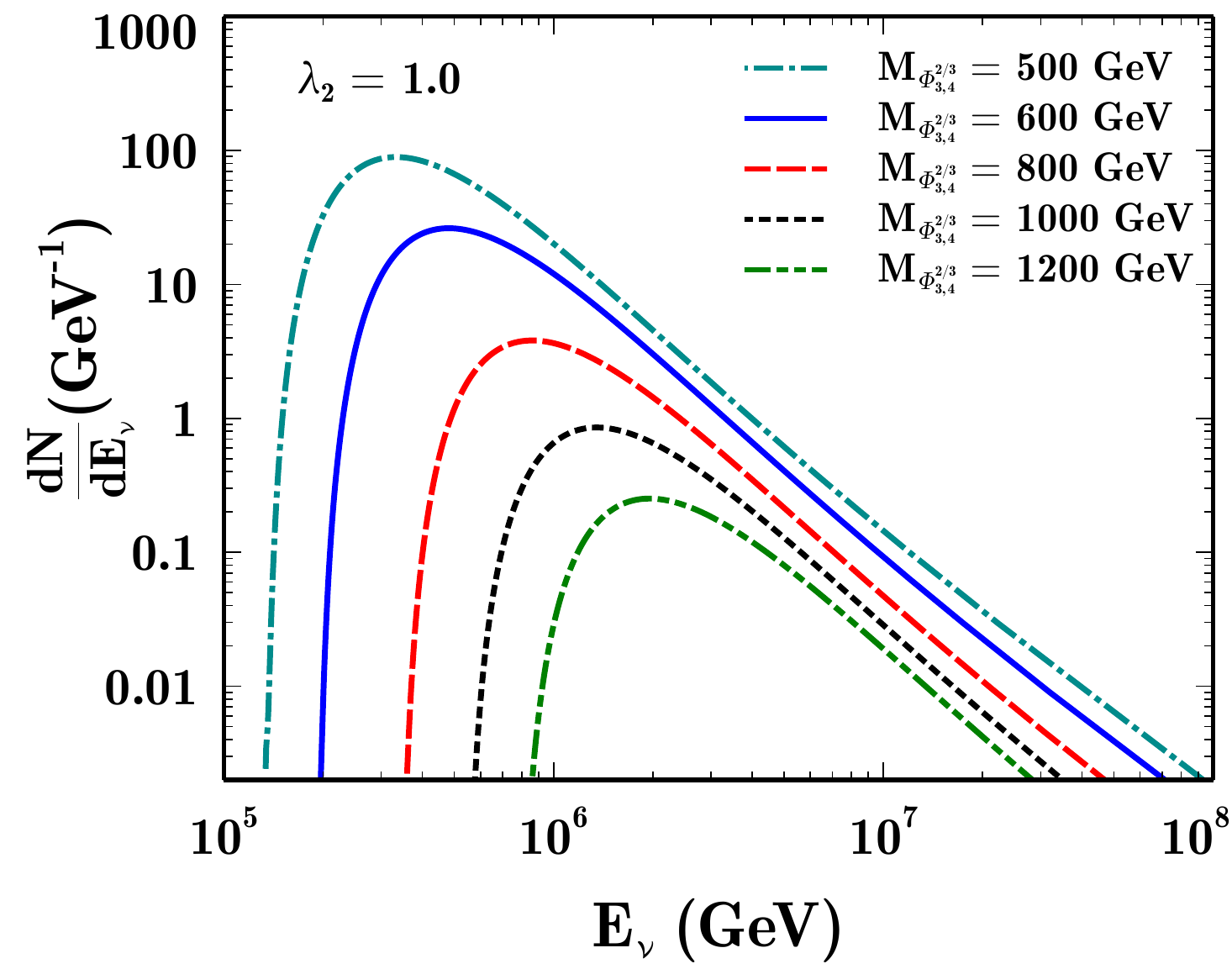}
}
\caption{The event rate distribution $dN/dE_{\nu}$ for various cases. The coupling $\lambda_{2}$ is shown in each plot.}
\label{f:dNdE}
\end{figure}

The event rate distributions, $\frac{dN}{dE_{\nu}}$,  are shown in Fig.~\ref{f:dNdE}. It is worth-mentioning that the rapidly falling nature of these distributions are dictated by the spectrum of neutrino flux $\Phi(E_{\nu})$.

\begin{figure}[!htbp]
\begin{center}
\includegraphics[scale=0.28,angle=-90]{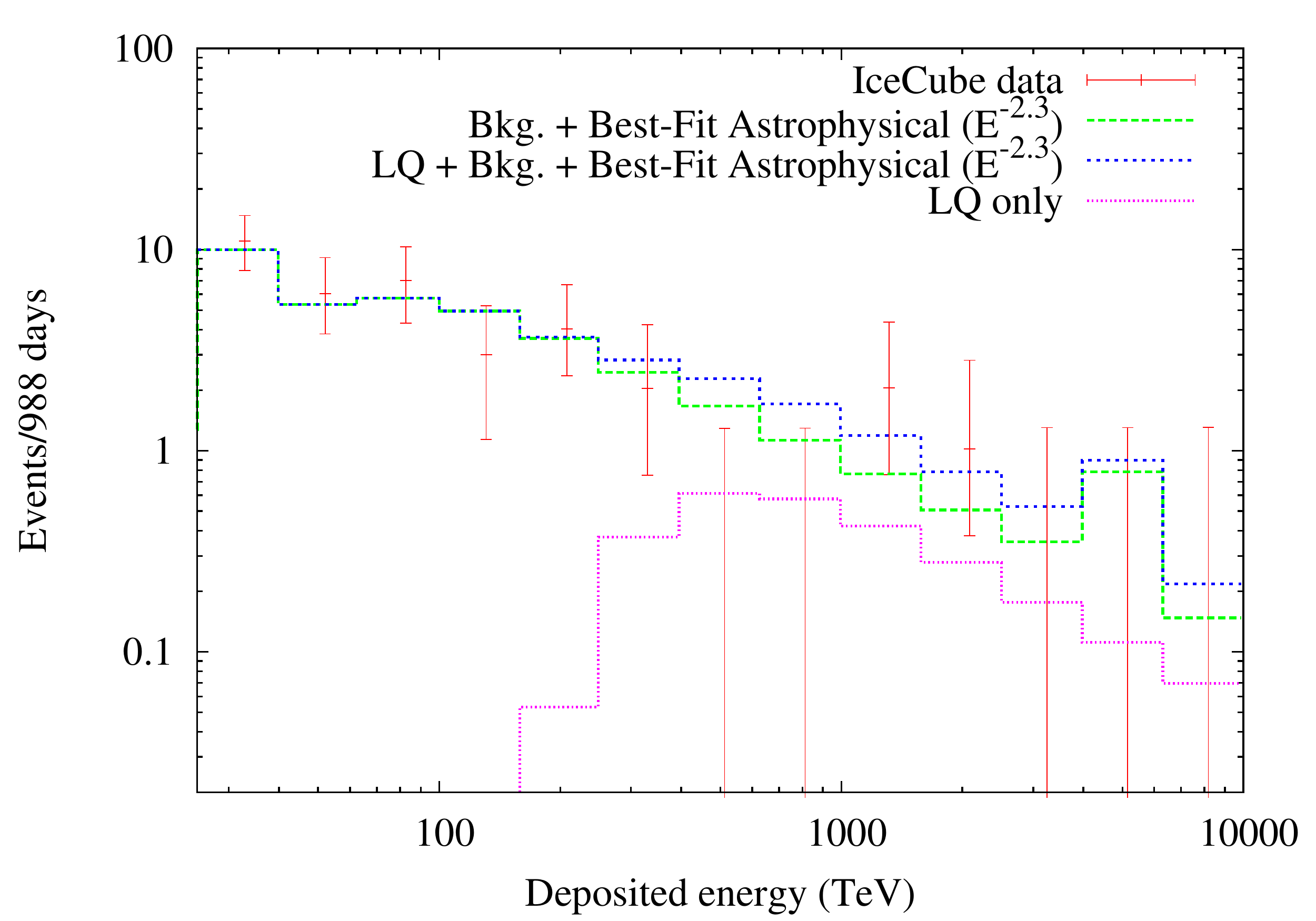}~~~
\includegraphics[scale=0.28,angle=-90]{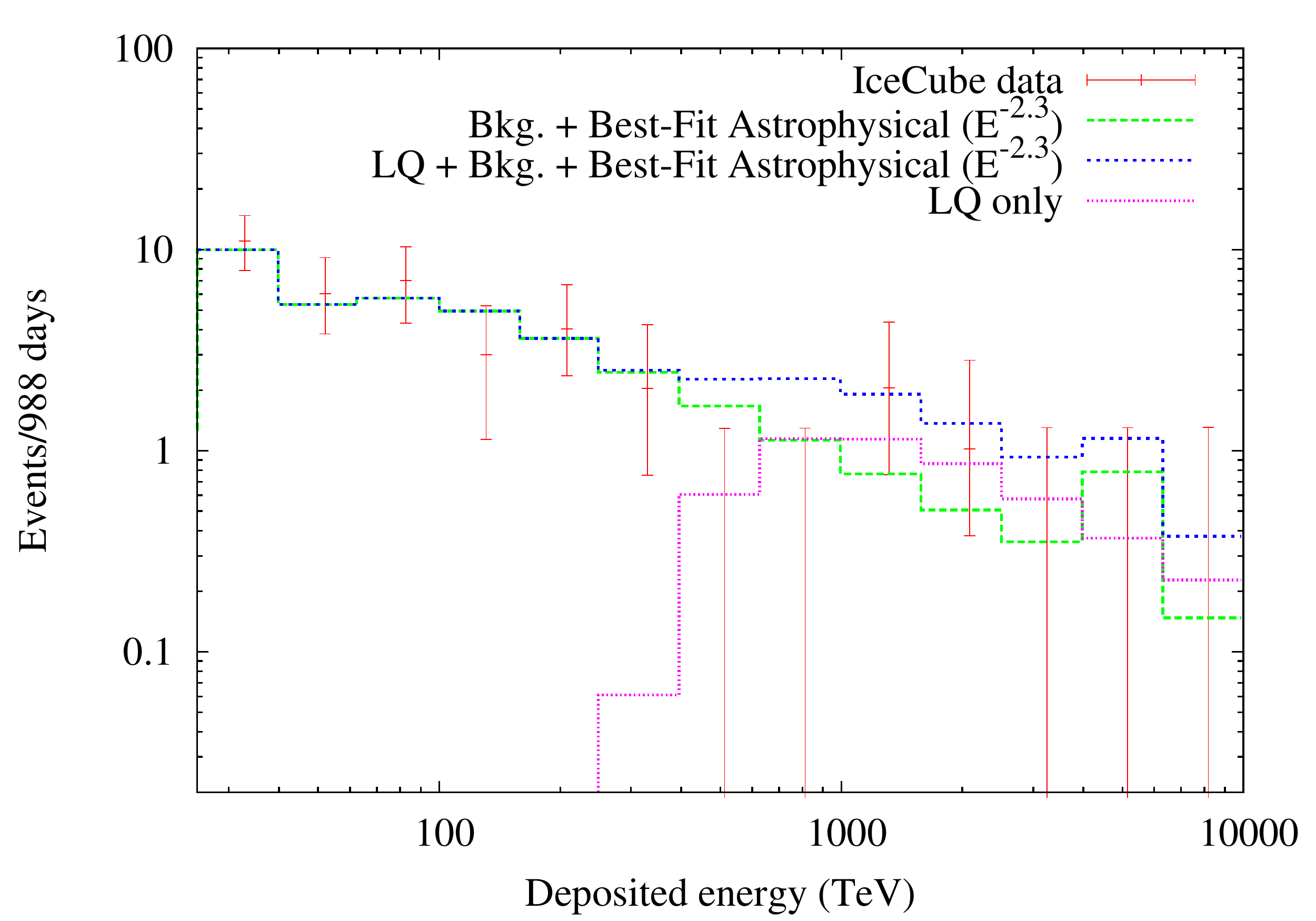}
\caption{Event rate distribution for  leptoquark with $\lambda_{2} = 1$ and $M_{\Phi_{3}^{-1/3}} = 500$ GeV (left) 650 GeV (right).}
\label{f:masscoup1}
\end{center}
\end{figure}

To test the allowed parameter space  of the $\lambda_2$ coupling and LQ mass for extra LQ events in the IceCube observations we do a $\chi$-square fit with $(\lambda_2, M_{\Phi_{3}^{-1/3}})$ as free parameters\footnote{Similar studies can be made for $\Phi_{3,4}^{2/3}$ also.}.
We define,  
\begin{align}
\chi^2(\lambda_2, M_{\Phi_{3}^{-1/3}}) =\sum_{\rm energy~bins}  \frac{ \left( \mathcal{N}_{exp} - (\mathcal{N}_{SM} +\mathcal{ N}_{LQ}) \right)^2 }{\Delta^2},
\end{align}
where the error $\Delta$ is the experimental error \cite{Aartsen:2014gkd} in the measured events in each energy bin and the uncertainty in the  model prediction added in quadrature. The events expected from the standard model CC and NC interactions 
from atmospheric neutrinos plus the extra-terrestrial neutrinos with flux spectrum (\ref{flux}) have been computed in \cite{Aartsen:2014gkd}. We add to the SM events the events expected from LQ processes and minimize the $\chi^2$ summing over energy bins from $E_{\nu}= 100~{\rm GeV}- 10~{\rm PeV}$.  The minimum $\chi^2$ is achieved for $\lambda_2=0$ which means that $LQ$ events spoil the overall fit from SM events. When the LQ mass is $M_{\Phi_{3}^{-1/3}}<500$ GeV then it gives an excess  contribution over observations in the $E_\nu < 1$ PeV energy bins where the fit from  standard model  CC and NC events from atmospheric neutrinos and  extra-terrestrial neutrinos with spectrum (\ref{flux}) fits observed IceCube events well. When the LQ mass is $M_{\Phi_{3}^{-1/3}}>650$ GeV (and the corresponding $\lambda_2 >1$, in order to maintain the cross section) there are excess events in the $E_\nu \sim 6$ PeV Glashow-resonance region where there is already a problem in that, even the SM prediction is not observed. In Fig.~\ref{f:masscoup1} we show the contribution of LQ with mass $M_{\Phi_{3}^{-1/3}}=500$ GeV and $M_{\Phi_{3}^{-1/3}}=650$ GeV respectively to the IceCube events. We see that in the mass range $M_{\Phi_{3}^{-1/3}}=500-650$ GeV with coupling $\lambda_2=1$ there is an improvement in the fit for the events in the $1-2$ PeV bins compared to the standard model.

\begin{figure}[!htbp]
\begin{center}
\includegraphics[height=4cm, width=5.5cm]{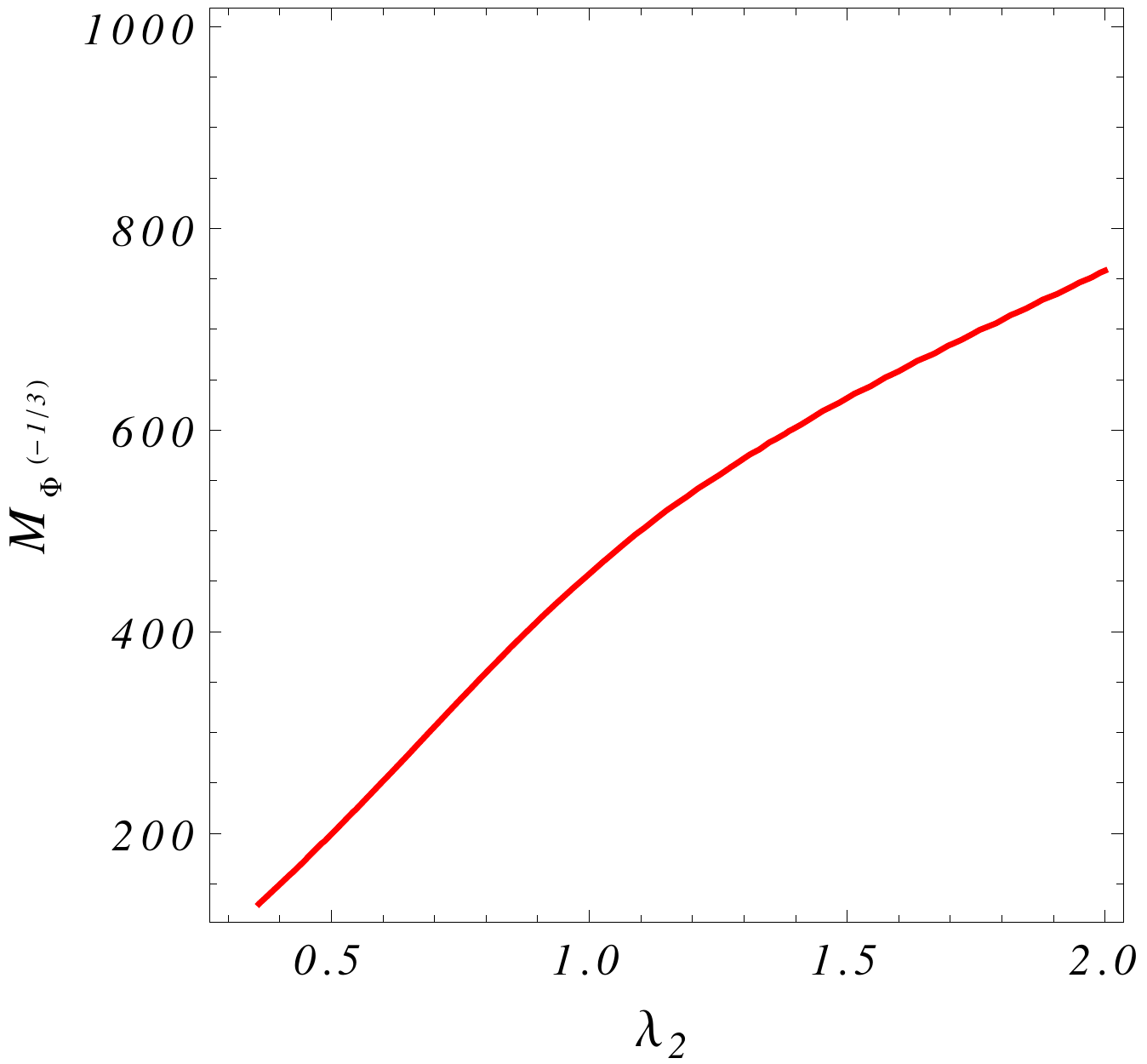}
\caption{The $(\lambda_2, M_{\Phi_{3}^{-1/3}})$ parameter space where the region below the curve is disallowed at 95\% CL. The $ \chi^{2}_{\rm min} = 3.66$  for $\lambda_2=0$ which means that when all the energy bins are considered in the $\chi^2$ the SM gives the best fit to the observed data.}
\label{f:masscoup3}
\end{center}
\end{figure}

As the total number spectrum of events has a worse $\chi^2$ fit with the inclusion of leptoquark events, we can rule out some parameter space of LQ models from IceCube events. In Fig.~\ref{f:masscoup3} we show the $(\lambda_2, M_{\Phi_{3}^{-1/3}})$ parameter space where the region below the curve is disallowed at 95\% CL. 


\section{Summary and Conclusion}
Leptoquarks in the mass range 500-1000 TeV can be produced in significant numbers by the PeV neutrinos at IceCube. Assuming that the 37 events seen in the IceCube data can be explained by $\nu_e,\nu_\mu,\nu_\tau$  CC and NC events we put bounds on the masses and couplings of scalar leptoquarks of the low scale quark-lepton unification model based on the gauge group $SU(4)_C \otimes SU(2)_L\otimes U(1)_R$~\cite{Perez:2013osa}. Of the possible signals in this model the $\bar \nu  q \rightarrow \bar \nu q$ process which will give rise to a hadronic shower that can be observed at IceCube. The $\bar \nu  q \rightarrow l q$ process allowed in the model which would produce lepton and hadronic showers is constrained to be small from collider bounds on the corresponding coupling. We put bounds on the scalar leptoquark masses which connect the first generation quarks with the third generation leptons to be above the $500$-$1000$ GeV if the coupling $\lambda_2 $ is in the $0.1$-$1$ range. Observations with longer exposures over target volumes can produce hadronic showers by resonant leptoquark decay which may be observed over the background. 

In the context of colliders, we would like to add that in this model the $eejj$ or $e\nu jj$ signal, studied in~\cite{Allanach:2015ria,Dutta:2015dka,Evans:2015ita}, will take place only via coupling $\lambda_{4}$ (see Eq.~\ref{e:lagrangian}), but this coupling is restricted by rare meson decay processes, which constrain $\lambda_{4}<0.01$~\cite{Smirnov:2007hv}. For these values of $\lambda_{4}$ the leptoquark events at IceCube are negligible compared to those generated by $\lambda_{2}$ coupling. This model does not explain the $eejj$ or $e\nu jj$ signals at LHC and is only constrained by IceCube data. Also due to the very nature of the couplings, in this model, one can have $\nu  \nu jj$ type of signal but that will be difficult to examine at colliders.

\section*{Acknowledgment}
We thank Soumya Rao, Aritra Gupta, Atri Bhattacharya and Abhaya Kumar Swain for valuable discussions.


\bibliographystyle{JHEP}
\bibliography{ref.bib}

\end{document}